\documentclass[preprint,showpacs,preprintnumbers,amsmath,amssymb]{revtex4}

\usepackage{graphicx}% Include figure files
\usepackage{dcolumn}% Align table columns on decimal point
\usepackage{bm}% bold math
\usepackage{amsmath}

\begin{document}

\preprint{}

\title{Ultra Fast Nonlinear Optical Tuning of Photonic Crystal Cavities}% Force line breaks with \\

\author{Ilya Fushman}
\email{ifushman@stanford.edu}
 \altaffiliation{Department of Applied Physics, Stanford University}%Lines break automatically or can be forced
\author{Edo Waks}
 \altaffiliation{Department of Electrical and Computer Engineering, 
University of Maryland, College Park, MD, 20742}%Lines break automatically or can be forced

\author{Dirk Englund}
\altaffiliation{Department of Applied Physics, Stanford University}%Lines break automatically or can be 
\author{Nick Stoltz}
 \altaffiliation{Department of Electrical and Computer Engineering, University of California Santa Barbara, CA 93106}
\author{Pierre Petroff}
\altaffiliation{Department of Electrical and Computer Engineering, University of California Santa Barbara, CA 93106}
\author{Jelena Vu\v{c}kovi\'{c}}
\altaffiliation{Department of Electrical Engineering, Stanford University}%Lines break automatically or can be 
 \affiliation{E. L. Ginzton Laboratory, Stanford University, Stanford, CA, 94305}%Lines break automatically or can be forced

\date{\today}% It is always \today, today,

\begin{abstract}

We demonstrate fast (up to 20 GHz), low power (5 $\mu W$) modulation of photonic crystal (PC) cavities in GaAs containing InAs quantum dots. Rapid modulation through blue-shifting of the cavity resonance is achieved via free carrier injection by an above-band picosecond laser pulse.  Slow tuning by several linewidths due to laser-induced heating is also demonstrated.  
\end{abstract}

\maketitle

\section{\label{sec:intro}INTRODUCTION:}
Nonlinear optical switching in photonic networks is a promising approach for ultrafast low-power optical data processing and storage \cite{ref:Joannop_nonlin_PC}. In addition, optical data processing will be essential for optics-based quantum information processing systems. A number of elements of an all optical network have been proposed and demonstrated in Silicon photonic crystals \cite{ref:Notomi_flip_flop,ref:Notomi_Review_NTT}. Tuning of the photonic crystal lattice modes has also been demonstrated \cite{ref:Krauss_Tune_Waveguide,ref:Vos_Tune}. Here, we directly observe ultrafast ($\approx$ 20 GHz) nonlinear optical tuning of photonic crystal (PC) cavities containing quantum dots (QD). We perform the fast tuning via free carrier injection, which alters the cavity refractive index, and observe it directly in the time domain. Three material effects can be used to quickly alter the refractive index. The first is the index change due to free carrier (FC) generation, which is discussed in this work, and has been explored elsewhere  \cite{ref:Krauss_Tune_Waveguide}.  The cavity resonance shifts to shorter wavelengths due to the free-carrier effect. Switching via free-carrier generation is limited by the lifetime of free carriers and depends strongly on the material system and geometry of the device. In our case, the large surface area and small mode volume of the PC reduce the lifetime of free carriers in GaAs. Free carriers can alternatively be swept out of the cavity by applying a potential across the device \cite{ref:Intel_Raman}. The second effect that can be used to modify the refractive index is the Kerr effect, which is promising for a variety of other applications \cite{ref:Hochberg_Nonlinear, ref:Kerr_Paper} and, in principle, should result in modulation rates of $10^{15}-10^{16}$ Hz.  However, the free carrier effect is more easily achieved in the GaAs PC considered here. The third effect is thermal tuning (TT) via optical heating of the sample through absorption of the pump laser. This process is much slower than free carrier and Kerr effects and shifts the cavity resonance to longer wavelengths due to the temperature dependence of the refractive index.  The time scale for this process is on the order of microseconds. Here we consider these two processes for modulating cavity resonances, and focus on the higher-speed FC tuning. 

	Photonic crystal samples investigated in this study are grown by molecular beam epitaxy on a Si n-doped GaAs (100) substrate with a 0.1µm buffer layer.  The sample contains a 10 period distributed Bragg reflector (DBR) mirror consisting of alternating layers of AlAs/GaAs with thicknesses of
80.2/67.6 nm respectively.  A 918 nm sacrificial layer of Al0.8Ga0.2As is located above the DBR mirror.  The active region consists of a 150 nm thick GaAs region with a centered InGaAs/GaAs QD layer.  QDÕs self-assemble during epitaxy operating in the Stranski-Krastanov growth mode.  InGaAs islands are partially covered with GaAs and annealed before completely capping with GaAs.  This procedure blue shifts the QDÕs emission wavelengths \cite{ref:Petroff} towards the spectral region where Si-based detectors are more efficient.

\begin{figure}
    \includegraphics[height=1.7in]{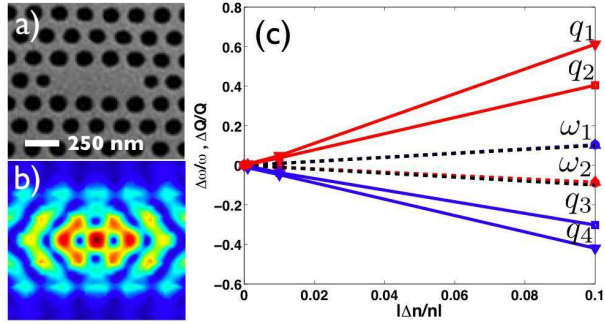}
   \caption{\textbf{(a):} Scanning electron micrograph of the L3 type cavity. \textbf{(b):} high-Q mode Electric field amplitude distribution, as predicted by FDTD simulations. \textbf{(c):} FDTD Simulations of frequency and Q changes as $\Delta n/n$ changes from $\pm10^{-3} \to \pm10^{-1}$. A high-Q ($Q_{HQ}=20000$) and low-L ($Q_{LQ}=2000$) cavity were tuned: \textbf{$(q_1)$} $\Delta Q/Q$ for $\Delta n > 0$ and $Q=Q_{HQ}$, \textbf{$(q_2)$}  $\Delta Q/Q$ for $\Delta n > 0$ and $Q=Q_{LQ}$, \textbf{$(\omega_1)$}  $\Delta \omega/\omega$ for $\Delta n < 0$,\textbf{$(\omega_2)$}  $\Delta \omega/\omega$ for $\Delta n > 0$ for both high Q and low Q modes, \textbf{$(q_3)$}  $\Delta Q/Q$ for $\Delta n < 0$  $Q=Q_{LQ}$, \textbf{$(q_4)$} $\Delta Q/Q$ for $\Delta n < 0$  $Q=Q_{HQ}$. The lines $\Delta n / n$ for $\Delta n > 0 $ and $\Delta n < 0$ are also plotted and overlap exactly with $\omega_2$ and $\omega_1$. As can be seen, the magnitude of the relative frequency change is independent of Q, but the higher Q cavity is degraded more strongly by the change in index. For an increase in n, the Q increases due to stronger Total Internal Reflection confinement in the slab, as expected. }
  \label{fig:fab-cav}
  \end{figure}

  PC cavities, such as those shown in Fig.\ref{fig:fab-cav}, were fabricated in GaAs membranes using standard electron beam lithography and reactive ion etching techniques. Finite Difference Time Domain (FDTD) simulations predict that the fundamental resonance in the cavity has a field maximum in the high index region (Fig.\ref{fig:fab-cav}), and thus a change in the value of the dielectric constant should affect these modes strongly. We investigated the dipole cavity (Fig.\ref{fig:fab-cav}), the linear three-hole defect cavity \cite{ref:NODAL3}, and the linear two-hole defect cavity designs. The experimentally observed Q's for all three cavities were in the range of 1000-2000 (optimized cavities can have much higher Q's), and consequently the experimental tuning results were similar for all three cavities.
  
	Photonic crystal cavities were made to spectrally overlap with the QD emission, and are visible above the QD emission background due to an increased emission rate and collection efficiency of dots coupled to the cavity. Quantum dot emission was excited with a Ti:Sapphire laser tuned to 750 nm in a pulsed or CW configuration. In the pulsed mode, the pump produced  3ps pulses at an 80 MHz repetition rate. Tuning was achieved by pulsing the cavity with appropriate pump power. The cavity emission was detected on a spectrometer and on a streak camera for the time resolved measurements.
  
     Tuning is achieved by quickly changing the value of the dielectric constant $\epsilon=n^2$ of the cavity with a control pulse. The magnitude of the refractive index shift $\Delta n$ can be estimated from 
\begin{equation}
\frac{\Delta\omega}{\omega} \approx -\frac{1}{2}\frac{\int \Delta \epsilon |E|^2 dV}{\int \epsilon |E|^2 dV} \approx -\frac{\Delta n}{n}
\end{equation}
Above, $\omega$ is the resonance of the un-shifted cavity, $|E|^2$ is the amplitude of the cavity mode, and the integral goes over all space. In order to shift by a linewidth, we require $\frac{\Delta \omega}{\omega}=\frac{1}{Q}$, which gives $\Delta n  = \frac{n}{Q}$. Finite difference time domain (FDTD) calculations indeed verify that for a linear cavity with Q ~ 1000, a $\Delta n \approx 10^{-2}$ shifts the resonance by more than a linewidth, as seen in Fig. \ref{fig:fab-cav}.

  As described above, two tuning mechanisms were investigated in this work. The first is temperature tuning, which is quite slow (on the time scale of microseconds). The second is the free carrier induced refractive index change, which is found to occur on the time scale of tens of picoseconds. Therefore, we can look at the two effects separately in the time domain. 
  
  \section{Free Carrier Tuning}
  
   In the case of Free Carrier (FC) tuning only,
  \begin{equation}
  \label{eq:dnn}
  \frac{\Delta n(t)}{n}=\frac{\Delta n_{fc}(t)}{n}=  \eta N_{fc}(t) 
   \end{equation}
 where $N_{fc}(t)$ is the density of free carriers in the GaAs slab, and the value of $\eta$ is given in terms of fundamental constants ($\epsilon_0$,c), DC refractive index ($n_0$), charge ($e$), effective electron mass  ($m^{*}_{e}$) and wavelength ($\lambda$) as $\eta=-\frac{e^2 \lambda^2}{8 \pi^2 c^2 \epsilon_0 n_0 m^{*}_{e}}$ \cite{ref:Nonlin_GaAs_Index} , and we calculate $\eta \approx 10^{-21} cm^{3}$ for our system. 
   
  The FC density changes with the pump photon number density $P(t)$, with pulse width $\tau_p$, in time t as:
  \begin{equation}
  \frac{d N_{fc}}{dt}=-\frac{1}{\tau_{fc}}N_{fc}+\frac{P(t)}{\tau_p}
  \end{equation}
  The carriers decay with $\frac{1}{\tau_{fc}}=\frac{1}{\tau_r}+\frac{1}{\tau_{nr}}+\frac{1}{\tau_c}$, where $\tau_r,\tau_{nr}$ are the radiative and non-radiative recombination times of free carriers, and $\tau_c$ is the relaxation time (or capture time) into the QDs. While $\tau_c \approx 30-50 ps << \tau_r,\tau_{nr}$, the dot capture is not the dominant relaxation process. The dots saturate for the duration of the dot recombination lifetime $\tau_d \approx 200ps-1ns$, and, because the dot density is much smaller than the FC density, the effective capture time is much longer. Qualitatively, we can describe this effect by lengthening $\tau_c$ by a factor 1/x as $\tau_c \to \tau_c/x << \tau_r,\tau_{nr}$, where $x<<1$ is essentially the ratio of QD to FC densities. The FC density is then given by: 
  \begin{equation}
  N_{fc}(t)=N_{fc}(0)e^{-\frac{t}{\tau_{fc}}}+e^{-\frac{t}{\tau_{fc}}}\int^{t}_{0}e^{\frac{t'}{\tau_{fc}}}\frac{P(t')}{\tau_p} dt'
  \end{equation} 
  
  In order to shift the cavity resonance by a linewidth ($\Delta n/n =10^{-3}$), we need $N_{fc} \approx 10^{18} cm^{-3}$ according to Eq. \ref{eq:dnn}. Taking into account the GaAs absorption coefficient $\alpha \approx 10^{4} cm^{-1}$, reflection losses from the 160 nm GaAs membrane ($R=(\frac{n_{air}-n_{GaAs}}{n_{air}+n_{GaAs}})^2 \approx .3$), lens losses ($50\%$), and an approximately $5 \mu m$ spot size, powers as low at 1-10 $\mu W$ average pulse power in a 3 ps pulse should yield the desired shifts of order $\frac{\Delta n}{n} \approx 10^{-3}$. 
  
    In our experiment, we monitor the cavity resonance during the tuning process using QD emission. Thus, we need to account for the delay between the pump and onset of emission in QDs. The QDs are excited by free carriers according to:
    \begin{equation}
  \frac{d N_{qd}}{dt}=-\frac{1}{\tau_{d}}N_{qd}+\frac{N_{fc}}{\tau_c}
  \end{equation}
  
    Thus the QD population (assuming no excited dots at carriers at t=0) is given by:
    \begin{equation}
  N_{qd}(t)=\frac{e^{-\frac{t}{\tau_{qd}}}}{\tau_p \tau_{c}}\int^{t}_{0}e^{t'(\frac{1}{\tau_{qd}}-\frac{1}{\tau_{fc}})}\int^{t'}_{0}e^{\frac{t''}{\tau_{fc}}}P(t'')dt'' dt'
  \end{equation} 
	where $\tau_p$ is the pump pulse width, $\tau_{qd} \approx 200$ ps is the average cavity coupled QD lifetime, $\tau_{fc} \approx 30 ps$ is the FC lifetime, and $P(t)$ is the pump photon number density.   The observed spectrum is that of a Lorentzian with a time - varying central frequency $\omega_0(t)$ (for simplicity, we assume that the Q  factor is time invariant), which we define as:
\begin{equation}
S(\omega,t)=(1+4 Q^2(1-\frac{\omega^{2}_0(t)}{\omega^{2}}))^{-1}
\end{equation}

\begin{figure}[!h]
  \includegraphics[height=2.75in]{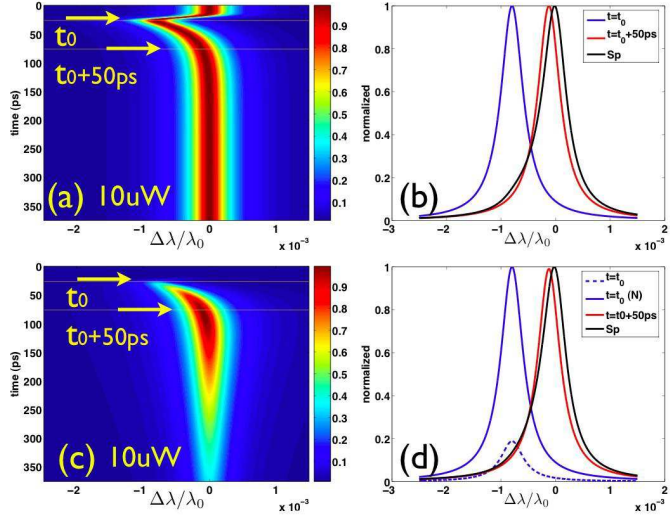}
  \caption{Numerical model of a free-carrier tuned cavity. In \textbf{(a)} the cavity is always illuminated by a light source. Panel \textbf{(b)} shows the cavity resonance at the peak of the free carrier distribution (t=0) and 50 ps later, as indicated by the yellow arrows in \textbf{(a)}. The time-integrated spectrum is shown as the asymmetric black line (lebeled Sp) in \textbf{(b)}, and corresponds to the signal seen on the spectrometer, which is the integral over the whole time window of the shifted cavity. The asymmetric spectrum indicates shifting. In \textbf{(c)} and \textbf{(d)} the same data is plotted, but now we consider the cavity illuminated only by QD emission with a turn-on delay of 30 ps due to the carrier capture lifetime $\tau_c$, and a QD lifetime of 200 ps. In \textbf{(d)} the dashed line is the un-normalized t=0 spectrum, which now appears much smaller in magnitude. Furthermore, the asymmetry of the line is even smaller in this case. 
  }
  \label{fig:fcmodel}
  \end{figure}

The numerical results are shown in Fig. \ref{fig:fcmodel}. We find that going beyond 10's of $\mu W$ does not result in a larger shift, but destroys and shifts the cavity Q permanently. 

\begin{figure}[!h]
  \includegraphics[height=2.75in]{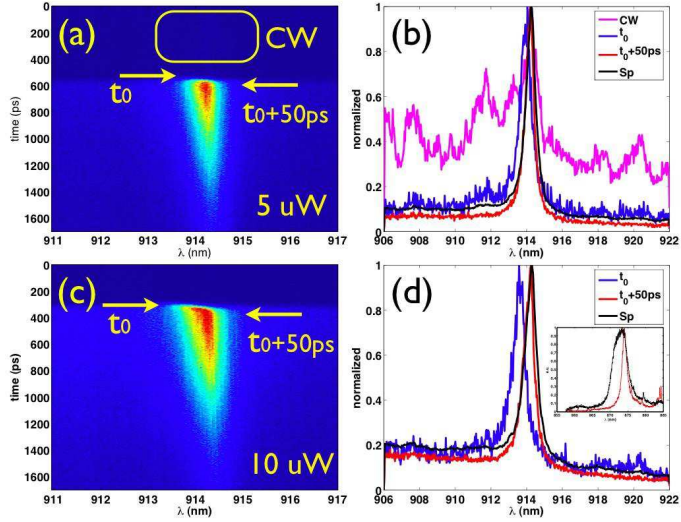}
  \caption{Experimental result of FC cavity tuning for the L3 cavity. In \textbf{(a)} the cavity is always illuminated by a light source and pulsed with a 3ps Ti:Sapphire pulse. Panel \textbf{(b)} shows the cavity resonance at the peak of the FC distribution (t=0) and 50 ps later, as indicated by the yellow arrows in \textbf{(a)}. In order to verify that the cavity tunes at the arrival at the pulse, we combine the pulsed excitation with a weak CW above band pump. The emission due to the CW source is always present, and is in the box labeled CW in \textbf{(a)}. This very weak emission is reproduced in Panel (b) as the broad background with a peak at the cold cavity resonace in \textbf{(b)}. The time-integrated spectrum is shown as the black line (Spectrometer) in \textbf{(b)}. In \textbf{(c)} and \textbf{(d)} the same data is plotted, but now we consider the cavity illuminated only by QD emission pulsed by 10 $\mu W$ from the Ti:Sapphire source. In \textbf{(d)} suppression by about .4-.35 at the cold cavity resonance can be seen.  The inset shows a strongly asymmetric spectrum of a dipole type cavity under excitation of $100 \mu W$ and the same cavity at low power after prolonged excitation. Such strong excitation degrades the Q. 
  }
  \label{fig:fcexperiment}
  \end{figure}

The experimental data is shown in Fig. \ref{fig:fcexperiment}. We used moderate power ($\approx$10 $\mu W$) to shift the cavity by one half linewidth. Stronger excitation results in higher shifts as indicated by an extremely asymmetric spectrum shown on the inset in (d) of Fig. \ref{fig:fcexperiment}, where 100 $\mu W$ were used. However, prolonged excitation at this power leads to a sharp reduction in Q over time.
      
 % \begin{figure}[!h]
  %\includegraphics[height=.75in]{streak-tune-rightparams}
  %\includegraphics[height=.75in]{power_overlay_good_crop}
  %\includegraphics[height=.75in]{streak-tune-fit-rightparams}
  %\caption{top: streak camera image of the free carrier induced dynamics. the red lines follow the center of the cavity resonance, which was obtained from the fit to a Lorentzian. bottom (inset): overlay of the narrow low power (red) and broad high power (black) spectrometer images. Such an asymmetric broad lineshape corresponds to a projection of the streak camera time-series onto a plane. }
  %\label{fig:pulse-tune}
  %\end{figure}

\section{Thermal Tuning}
In the case of Thermal Tuning (TT), 
\begin{equation}
  \frac{\Delta n(t)}{n}= \beta T 
 \end{equation}

\begin{figure}[!h]
 \includegraphics[height=1.75in]{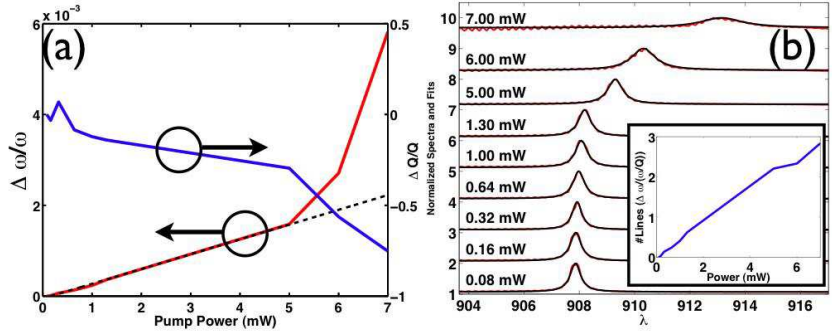}
  \caption{  Thermal tuning of the L3 cavity under CW excitation \textbf{(a)}: Measured $\Delta\omega/\omega$ (left axis) and  $\Delta Q/Q$ (right axis)  as a function of pump power for the L3 cavity, obtained from the fits to the spectra shown in \textbf{(b)}. The Q initially increases due to moderate gain and then degrades, while $\omega$ shifts linearly. The straight dashed line fits $\Delta\omega/\omega = 3\times10^{-3}\times P_{in}-5 \times 10^{-5}$ with $95\%$ confidence and with root mean square deviation of $ \approx 0.99$. At very high power, the change in frequency does not follow the same trend. The inset in \textbf{(b)} shows a plot of $\Delta\omega/(\omega/Q)$, which is a measure of the number of lines that we shift the cavity by. A shift of three linewidths is obtained.}
  \label{fig:cw-tune}
  \end{figure}

   Continuous  wave above-band excitation of the sample results in both free carrier generation and heating. In this case, the heating mechanism dominates, and the cavity red-shifts. The predominant effect on the dielectric constant is the change in the bandgap with temperature due to lattice expansion and phonon population. The cavity itself could potentially expand, but since the thermal expansion coefficient of GaAs is on the order of $10^{-6} K^{-1}$, this is insignificant. As the cavity red-shifts, the Q first increases due to gain and then drops due to absorption losses. The experimental data for thermal tuning is shown in Fig. \ref{fig:cw-tune}. From a fit to the frequency shift, we obtain $\beta \approx 3\times 10^{-3}$.

\section{\label{sec:conc} CONCLUSION:}
	In conclusion, we show that fast (20 GHz) tuning of GaAs cavities can be realized with reasonable pump powers (10 $\mu W$) with no additional fabrication. Under these conditions the cavity is shifted by almost a linewidth, which leads to suppression of transmission at the cold-cavity frequency by $\approx 1/e$. The suppression depends on the Q of the cavity and for cavities with Q $\approx 4000$, shifts by a full linewidth would be obtained. Thus, fast control over photon propagation in a GaAs based PC network is easily achieved and can be used to control the elements of an optical or quantum on-chip network. Free carrier tuning strongly depends on the geometry of the cavity, since a larger surface area leads to a shorter FC lifetime. Thus, our future work will focus on identifying optimal designs for shifting and  a demonstration of an active switch based on the combination of PC cavities and waveguides. Our ultimate goal is all-optical logic with photon packets on the chip.

%section{\label{sec:ack} Acknowledgements:}

Financial support was provided by the MURI Center for photonic quantum information systems (ARO/DTO program No. DAAD19-03-1-0199), ONR Young Investi-gator Award and NSF Grant No. CCF-0507295. I.F. and D.E. would like to thank the NDSEG fellowship for financial support. 

\bibliography{APL_submission_1_NonlinearTuning_ARXIV1}

\end{document}